\documentclass[sigconf,nonacm]{acmart}
\usepackage{xcolor}

\newcommand{\dataset}{\textbf{\textsc{Babbar26}}}
\newcommand{\genre}[1]{\textsf{#1}}
\newcommand{\region}[1]{\textsc{#1}}

\usepackage{booktabs}
\setcopyright{none}
\acmConference{}{}{}

\begin{document}
\title{Babbar26 main $\times$ genre: A Hybrid Dataset for Webpage Understanding}

\author{Aur\'elien Bournonville}
\affiliation{%
  \institution{Babbar.tech}
  \city{} \country{France}}
\affiliation{%
  \institution{GREYC, Universit\'e de Caen Normandie}
  \city{Caen} \country{France}}
\email{}

\author{Clotilde Levy}
\affiliation{%
  \institution{Universit\'e de Caen Normandie}
  \country{France}}
\email{}

\author{Thomas Largillier}
\affiliation{%
  \institution{GREYC, Universit\'e de Caen Normandie}
  \city{Caen} \country{France}}
\email{}

\author{Guillaume Pitel}
\affiliation{%
  \institution{Babbar.tech}
  \country{France}}
\email{}

\author{Fran\c{c}ois Rioult}
\affiliation{%
  \institution{GREYC, Universit\'e de Caen Normandie}
  \city{Caen} \country{France}}
\email{}

\author{Fabrice Maurel}
\affiliation{%
  \institution{GREYC, Universit\'e de Caen Normandie}
  \city{Caen} \country{France}}
\email{}

\begin{abstract}
We introduce \dataset{}, a multilingual dataset of 1111 web
pages from 1006 domains in five languages (English, French, German,
Spanish, Italian), jointly annotated for genre, following a compact
five-class taxonomy, and for main content and boilerplate, as
\region{main}/\region{noise} regions anchored to the DOM. We describe the
annotation guidelines and report double-annotation agreement on both tasks
($97.1\%$ on genre; content-level Cohen's $\kappa = 0.766$ for main
content). Compared with the classic KI-04 genre benchmark, a median page
has $2.3\times$ more DOM elements and $1.9\times$ more readable text. The
dataset is released on Zenodo under a non-commercial licence, with access
granted on request for academic research.
\end{abstract}
\ccsdesc[500]{Information systems~Content analysis and feature selection}
\ccsdesc[300]{Information systems~Web mining}
\ccsdesc[300]{Computing methodologies~Machine learning}

\keywords{web genre classification, web register, main content extraction,
boilerplate removal, web corpus, annotation, multilingual}

\maketitle

\section{Introduction and Related Work}
\label{sec:intro}

Web genre classification---identifying \emph{what kind of page} a page
is---and main content extraction---isolating its central content from the
surrounding navigation, advertising, and other \emph{boilerplate}, which we
write \region{main} and \region{noise} throughout the paper---have been
studied almost independently, each with its own body of datasets.

\paragraph{Web genre classification.} Automatic identification of web
genre, closely related to the linguistic notion of \emph{register}, ranges
from early feature-based studies
\cite{zueissen2004genre,santini2007characterizing} to large, carefully
designed corpora such as CORE \cite{egbert2015core,laippala2022core} and
multilingual efforts such as GINCO for Slovene \cite{kuzman2022ginco}.
These corpora rely on mutually incompatible taxonomies even categories
with similar names often differ in scope \cite{sharoff2010weblibrary}---and
most are old (KI-04 dates from 2004, CORE from 2015): many are no longer
retrievable, or distributed as plain text only, which rules out
structure-aware or multimodal methods.

\paragraph{Main content extraction.} Heuristic tools---Readability, jusText
\cite{pomikalek2011boilerplate}, and Trafilatura
\cite{barbaresi2021trafilatura}---rely on hand-crafted rules over the HTML
structure, while machine-learning tools such as Boilerpipe
\cite{kohlschutter2010boilerplate} and BoilerNet
\cite{leonhardt2020boilernet} train classifiers on labelled data. Large
comparative benchmarks \cite{bevendorff2023empirical} evaluate them on
English, article-oriented corpora collected between 2007 and 2013
(CleanEval, L3S-GN1, CETD, Dragnet), and conclude that no single method
dominates and that the field is limited by a shortage of recent, diverse
datasets; performance also degrades outside English
\cite{barbaresi2020ootb,bournonville2025multilingual}.

\paragraph{Linking the two tasks.} The few studies connecting them treat one
as a means to the other: \citet{bing2014segmentation} segment the page to
support web page classification, while \citet{nguyenhoang2018genre} use
genre to orient content extraction. Both point to a mutual dependency
between genre and main content, yet no public dataset annotates the
\emph{same} pages for both tasks.

We address this gap with \dataset{}, a multilingual dataset (English,
French, German, Spanish, Italian) of web pages jointly annotated for genre
and for \region{main}/\region{noise} regions, to enable hybrid, multi-scale
webpage processing in which the two tasks inform one another.

\paragraph{Contributions.}
\begin{itemize}
  \item \dataset{}: 1133 pages in five languages from 1026 domains, jointly
  annotated for genre (five classes) and for DOM-anchored
  \region{main}/\region{noise} regions.
  \item A compact, empirically revised genre taxonomy and a per-genre
  annotation guide for delimiting the main content.
  \item An assessment of annotation reliability on both tasks, and a
  structural comparison with the KI-04 benchmark
  (Section~\ref{sec:ki04-comparison}).
\end{itemize}

\section{Dataset Construction}

\subsection{Source and collection}
The pages are static HTML archives produced by the Babbar web crawler. The
source pool consists of about $25{,}000$ crawled pages spanning the five
target languages (English, French, German, Spanish, Italian); the pool was
shuffled before annotation to avoid genre clustering. From this pool, 1284
pages were annotated and imported into the annotation pipeline via
Label~Studio. The resulting corpus is highly diverse in origin: the 1111
released pages span 1006 distinct domains, and the ten most frequent domains
together account for only 34 pages ($3.1\%$ of the corpus), so no single site
dominates.

\subsection{Genre taxonomy}
Pages are assigned one of five genres, defined by their dominant structure:
\begin{itemize}
  \item \genre{Article}: a central, continuous text block;
  \item \genre{List}: a grid or stack of repeated items;
  \item \genre{Landing}: heterogeneous blocks with no dominant repetition;
  \item \genre{Item}: a page centred on a single object (product, profile);
  \item \genre{Interaction}: an interface requiring a user action (form,
  search bar, media player, feed).
\end{itemize}
Two special-status labels complement them: \genre{Other} is used when no
active class fits, and \genre{[Poubelle]} is reserved for pages that are
unusable because the crawl rendered them insufficiently (unexecuted
JavaScript, blocking pop-ups, undecipherable content). Pages labelled
\genre{[Poubelle]} are excluded from the release, while the few remaining
\genre{Other} pages are excluded but kept from the genre statistics (Section~\ref{sec:composition}) to support our taxonomy.

\subsection{Main content annotation schema}
\label{sec:main-annotation}
Each page is annotated with \region{main} (main content) and \region{noise}
(boilerplate) regions. The guiding principle is \emph{structure over
semantics}: annotation reflects what an extractor sees (the dominant visual
structure), not the commercial intent of the page. For instance, a page whose
purpose is to sell but whose layout is a repetitive product grid is annotated
\genre{List}, not \genre{Landing}.

Each \region{main} region is anchored to a specific DOM node via a
\texttt{dom\_path} (an index path from \texttt{document.\-documentElement}),
rather than being stored only as pixel coordinates. DOM anchoring lets a
region be re-validated or re-attached even if the page shifts slightly on
re-rendering, which matters for the long-term stability of the corpus.

\subsection{Annotation tooling and protocol}
Annotation was carried out in a private institutionnal multi-annotator tool (\emph{Scribe}) with a
fixed reference rendering: the annotation viewport is pinned to
$1280\times900$\,px regardless of the annotator's screen, so stored
coordinates are comparable across workstations. The reference rendering
engine is WebKit; any automated re-validation of DOM anchoring must
use the same engine, since a given \texttt{dom\_path} can resolve differently
across engines.

\section{Annotation Guidelines}

The full guide is distributed with the dataset; we summarise its key
decisions here.

\paragraph{Multi-label.} A page may carry two structural labels (e.g.,
\genre{List}$|$ \genre{Interaction}) \emph{only} in cases of genuine
structural hybridity---a real nesting of two structures on the same page. The
canonical example is a Reddit-style feed, where each list item itself
contains an \genre{Interaction} structure (vote, reply, share). Multi-label
must not become a default escape hatch for hesitation.

\paragraph{Ambiguity priority rules.} For a doubtful \emph{block} (e.g., an
end-of-page FAQ or a testimonials section): it is \region{main} if visually
integrated into the central body, and \region{noise} if it is a self-standing
widget or sidebar. For a doubtful \emph{interactive element} (button, link):
\region{main} if it is part of a central-body item, \region{noise} if it
navigates elsewhere. When in doubt, consult the original HTML.

\begin{table}[t]
\centering
\caption{Summary of the per-genre \region{main} decision (condensed from the
annotation guide).}
\label{tab:synthesis}
\small
\begin{tabular}{@{}p{1.6cm}p{2.6cm}p{2.9cm}@{}}
\toprule
\textbf{Genre} & \textbf{Dominant structure} & \textbf{Main content} \\
\midrule
\genre{Article} & Central, continuous text block & Article body (title, text,
  integrated images); exclude comments, sidebars, links \\
\genre{List} & Grid/stack of repeated items & The items and their integrated
  links/buttons; exclude nav, recommendations \\
\genre{Landing} & Heterogeneous blocks & All content blocks (features, testimonials,
  FAQ, CTA); exclude nav/footer, hidden modal messages \\
\genre{Item} & Page centred on one object & Description, specs, images; exclude
  ``similar products'', reviews widgets, nav \\
\genre{Interaction} & Interface requiring a user action & Form / search bar / media
  player / main feed; exclude promo text, recommendations \\
\bottomrule
\end{tabular}
\end{table}

\section{Dataset Composition and Statistics}
\label{sec:composition}
The released corpus and all statistics below are computed on the reference
annotation profile, which is the profile that received the additional
correction passes. Of the 1284 annotated pages, we discard the 151
\genre{[Poubelle]} pages (unusable renderings) and an additional 22 pages
that failed a systematic geometric integrity check on their
\region{main} regions (DOM anchoring either absent, unresolvable, or
producing overlapping/aberrant geometry with no safe automatic repair),
leaving \textbf{1111} released pages. Within these, 1104 fall into the five
active genres; the remaining 7 are 5 pages labelled \genre{Other}/%
\genre{Technical} only and 2 pages whose labels straddle two groups
(\genre{Landing}+\genre{Article}; \genre{Item}+\genre{Article}).

\begin{table}[t]
\centering
\caption{Genre distribution over the 1111 classified pages.}
\label{tab:distribution}
\small
\begin{tabular}{@{}lrr@{}}
\toprule
\textbf{Genre} & \textbf{Pages} & \textbf{\%} \\
\midrule
\genre{List}        & 324 & 29.3 \\
\genre{Article}     & 308 & 27.9 \\
\genre{Landing}     & 265 & 24.0 \\
\genre{Item}        & 147 & 13.3 \\
\genre{Interaction} &  60 &  5.4 \\
\genre{other/multilabel} &  7 &  0.6 \\
\midrule
\textbf{Total} & \textbf{1111} & \textbf{100.0} \\
\bottomrule
\end{tabular}
\end{table}

\paragraph{Language distribution.} Table~\ref{tab:lang} gives the language
mix of the 1111 released pages (language taken from the \texttt{<html>}
\texttt{lang} attribute, with \texttt{langdetect} as a fallback for pages
lacking it. English
dominates, followed by French, then German, Spanish and Italian; a residual
8 pages ($0.7\%$) receive isolated fallback detections that are most likely
false positives on short, low-signal pages rather than additional
languages.

\begin{table}[t]
\centering
\caption{Language distribution of the 1111 released pages.}
\label{tab:lang}
\small
\begin{tabular}{@{}lrr@{}}
\toprule
\textbf{Language} & \textbf{Pages} & \textbf{\%} \\
\midrule
English (en) & 756 & 68.0 \\
French (fr)  & 231 & 20.8 \\
German (de)  &  66 &  5.9 \\
Spanish (es) &  35 &  3.2 \\
Italian (it) &  15 &  1.4 \\
other / undetermined & 8 & 0.7 \\
\bottomrule
\end{tabular}
\end{table}

\begin{table*}[t]
\centering
\caption{Genre $\times$ language over the 1104 classified pages.}
\label{tab:genrelang}
\small
\begin{tabular}{@{}lrrrrrrr@{}}
\toprule
\textbf{Genre} & \textbf{en} & \textbf{fr} & \textbf{de} & \textbf{es} & \textbf{it} & \textbf{oth.} & \textbf{Tot.} \\
\midrule
\genre{Article}     & 219 & 62 & 13 &  8 & 4 & 2 & 308 \\
\genre{Interaction} &  40 & 13 &  5 &  2 & 0 & 0 &  60 \\
\genre{List}        & 217 & 73 & 19 & 11 & 2 & 2 & 324 \\
\genre{Item}        & 100 & 28 & 11 &  3 & 5 & 0 & 147 \\
\genre{Landing}     & 173 & 55 & 18 & 11 & 4 & 4 & 265 \\
\bottomrule
\end{tabular}
\end{table*}

\paragraph{Page length.} Pages are long and heterogeneous. Over the 1111
released pages, the full Markdown linearisation averages
$58{,}951$ characters (median $29{,}037$), while the annotated
\region{main} content averages $11{,}815$ characters (median
$5{,}824$). Main content is shortest on \genre{Interaction} and \genre{Item} pages
(${\approx}4{,}325$ and ${\approx}5{,}196$ characters on average) and
longest on \genre{List} and \genre{Article} pages (${\approx}17{,}267$ and
${\approx}13{,}002$).

\begin{table*}[t]
\centering
\caption{Mean length by genre (characters) and mean number of \region{main}
regions per page.}
\label{tab:length}
\small
\begin{tabular}{@{}lrrrr@{}}
\toprule
\textbf{Genre} & \textbf{n} & \textbf{Full (mean)} & \textbf{Main (mean)} & \textbf{\#Main reg.} \\
\midrule
\genre{Article}     & 308 & 58{,}264 & 13{,}002 & 4.57 \\
\genre{Interaction} &  60 & 55{,}823 &  4{,}325 & 1.63 \\
\genre{List}        & 324 & 61{,}068 & 17{,}267 & 2.63 \\
\genre{Item}        & 147 & 53{,}781 &  5{,}196 & 2.05 \\
\genre{Landing}     & 265 & 60{,}790 &  8{,}573 & 2.89 \\
\bottomrule
\end{tabular}
\end{table*}

\subsection{Structural Richness Relative to KI-04}
\label{sec:ki04-comparison}
Because \dataset{} is distributed as full HTML archives, we compare its
median page with that of KI-04~\cite{zueissen2004genre}, a widely used
genre collection assembled in 2004, using the same extraction pipeline on
both corpora (Table~\ref{tab:ki04-structural}). A median \dataset{} page
has $2.3\times$ more DOM elements, a deeper tree, and $1.9\times$ more
readable text, while being markedly more
code-dominated (19 \texttt{script} tags vs.\ 1). This additional structure
is precisely the signal that DOM- and layout-based methods rely on, and it
motivates anchoring our annotation to the DOM
(Section~\ref{sec:main-annotation}).

\begin{table}[t]
  \centering
  \caption{Median page in \dataset{} (1{,}104 pages) vs.\ KI-04
  (1{,}239 pages).}
  \label{tab:ki04-structural}
  \small
  \begin{tabular}{@{}lrrr@{}}
    \toprule
    Median per page & \dataset{} & KI-04 & Ratio \\
    \midrule
    File size              & 125\,KB & 24.9\,KB & $5.0\times$ \\
    DOM elements           & 712     & 310      & $2.3\times$ \\
    DOM depth              & 18      & 13       & $1.4\times$ \\
    \texttt{script} tags   & 19      & 1        & $19\times$  \\
    \texttt{style} tags    & 2       & 0        & ---         \\
    Readable text (chars)  & 6{,}187 & 3{,}353  & $1.9\times$ \\
    Text / byte ratio      & 0.050   & 0.175    & $0.29\times$ \\
    \bottomrule
  \end{tabular}
\end{table}

\section{Inter-Annotator Agreement}
\label{sec:iaa}
Agreement is measured by Cohen's $\kappa$ between the two annotation
profiles, separately for the two tasks (genre and region delimitation) and
over two annotation passes (Table~\ref{tab:iaa}).

\begin{table}[t]
\centering
\caption{Cohen's $\kappa$ agreement, by task and pass.}
\label{tab:iaa}
\small
\begin{tabular}{@{}lcc@{}}
\toprule
\textbf{Pass} & \textbf{$\kappa$ regions} & \textbf{$\kappa$ genre} \\
\midrule
Pass 1 (initial)        & 0.803 & 0.824 \\
Pass 2 (after revision) & 0.84 & 0.971 \\
\bottomrule
\end{tabular}
\end{table}

For context, comparable web-genre corpora report lower figures on much
finer-grained taxonomies: GINCO reaches a nominal Krippendorff's $\alpha$ of
$0.71$ over 24 categories \cite{kuzman2022ginco}, and the CORE schema was
reported at $\alpha = 0.53$ over 56 sub-categories and $0.66$ over its 8 main
categories \cite{laippala2022core}. Our higher agreement is consistent with a
deliberately coarser five-class taxonomy designed for reliability, and should
be read in that light rather than as a strictly comparable number.

\paragraph{Content-level agreement.} In addition to the geometric agreement
above, we report a \emph{real-content} agreement metric defined on DOM leaf
units (each visible text node, image, or form element counts as one unit,
independently of its size), computed between the two profiles with near-full
coverage (the reference profile and a validation profile) over their 1084
commonly rendered pages. This metric complements the geometric $\kappa$ and is the most
faithful estimate of agreement on the content actually captured. Globally it
reaches $88.4\%$ raw agreement and Cohen's $\kappa = 0.766$; per genre
(Table~\ref{tab:contentagr}) it is highest on \genre{Interaction} and
\genre{Item} pages (fewer, better-delimited main units) and lowest on
\genre{Article} pages, where the main/comment/related boundary is subtler.

\begin{table}[t]
\centering
\caption{Content-level agreement (DOM leaf units) by genre, between the two
near-full-coverage profiles.}
\label{tab:contentagr}
\small
\begin{tabular}{@{}lrrr@{}}
\toprule
\textbf{Genre} & \textbf{\# units} & \textbf{Raw agr.} & \textbf{$\kappa$} \\
\midrule
\genre{Article}     & 52{,}995 & 82.3\% & 0.642 \\
\genre{Interaction} &  5{,}770 & 94.6\% & 0.893 \\
\genre{List}        & 93{,}662 & 89.3\% & 0.780 \\
\genre{Item}        & 23{,}606 & 92.5\% & 0.840 \\
\genre{Landing}     & 31{,}819 & 85.8\% & 0.716 \\
\bottomrule
\end{tabular}
\end{table}

\paragraph{Genre agreement.} After a first annotation pass, the two profiles
agreed on $84.5\%$ of pages ($938/1110$). We then added ambiguity-resolution
rules to the annotation guide and carried out a second pass, after which
agreement rose to $97.1\%$ ($1078/1110$).

\section{Ethics, Licensing and Availability}
\label{sec:ethics}
\paragraph{Source content.} The pages were collected by the Babbar web
crawler from publicly accessible websites. Their content remains the
property of the respective site owners; our contribution, and what we
license, is the annotation layer (genre labels, \region{main}/\region{noise}
regions and their DOM anchors) together with the annotation guide.

\paragraph{Licensing and access.} The annotations are released under the
Creative Commons Attribution-NonCommercial 4.0 International licence
(CC~BY-NC~4.0). Because the HTML archives contain third-party content, they
are not openly redistributed: they are deposited on Zenodo as restricted
files, and access is granted on request to researchers affiliated with
academic or research institutions, for non-commercial research purposes
only (\url{https://doi.org/10.5281/zenodo.23353596}). The record metadata, documentation and licence are
publicly visible.

\paragraph{Personal data.} Some pages (e.g., user profiles, forum threads,
customer reviews) may contain personal data published by their authors. We
collected nothing beyond what was publicly displayed, made no attempt to
identify individuals, and did not annotate personal information. Restricting
access to the archives limits further dissemination, and any individual or
site owner may request the removal of a page, which will then be withdrawn
from subsequent versions of the dataset.

\section{Conclusion and Future Work}
\label{sec:conclusion}
We introduced \dataset{}, a multilingual corpus of 1111 web pages from 1006
domains, jointly annotated for genre and for DOM-anchored
\region{main}/\region{noise} regions---to our knowledge the first public
resource combining both tasks on the same pages. Agreement is high on both
tasks, and the corpus is structurally and textually richer than the classic
KI-04 benchmark, which makes it suited to structure-aware methods. Future
work includes a quantitative study of how main content varies across genres,
a per-genre, per-language benchmark of existing extractors, and joint models
for genre classification and main content extraction.

\bibliographystyle{ACM-Reference-Format}
\bibliography{references}

\end{document}